\documentclass[twocolumn,showpacs,preprintnumbers,amsmath,amssymb,amsmath,amssymb,prb,floatfix,lengthcheck,natbib]{revtex4-1}

\usepackage{graphicx}
\usepackage{bm}
\usepackage{hyperref}
\usepackage{color}
\usepackage{natbib}

\begin{document}

\title{AFM probe for the signatures of Wigner correlations in the conductance of a one-dimensional quantum dot}
\author{N. Traverso Ziani$^{1,2}$, F. Cavaliere$^{1,2}$ and M. Sassetti$^{1,2}$}
 \affiliation{$^1$ Dipartimento di Fisica, Universit\`a di Genova, Via Dodecaneso 33,
  16146, Genova, Italy.\\
 $^2$ CNR-SPIN, Via Dodecaneso 33,
  16146, Genova, Italy.}
\date{\today}

\begin{abstract}
The transport properties of an interacting one-dimensional quantum dot
capacitively coupled to an atomic force microscope probe are
investigated. The dot is described within a Luttinger liquid framework
which captures both Friedel and Wigner oscillations. In the linear
regime, we demonstrate that both the conductance peak position and
height oscillate as the tip is scanned along the dot. A pronounced
beating pattern in the conductance maximum is observed, connected to
the oscillations of the electron density. Signatures of the effects
induced by a Wigner molecule are clearly identified and their
stability against the strength of Coulomb interactions are
analyzed. While the oscillations of the peak position due to Wigner
get enhanced at strong interactions, the peak height modulations are
suppressed as interactions grow. Oscillations due to Friedel, on the
other hand, are robust against interaction.
\end{abstract}

\maketitle

\section{Introduction}
\label{sec:intro}
Quantum dots~\cite{koudots} are an ideal playground to study the
interplay between quantum confinement and Coulomb interactions. Beyond
the well-known Coulomb blockade physics, lie several interesting
physical effects which induce peculiar correlations among electron
states.\\
\noindent In two-dimensional (2D) quantum dots such correlations have
been the subject of an intense theoretical research, especially
employing several different numerical
techniques.~\cite{2Dnum1,2Dnum2,2Dnum3,2Dnum4,2Dnum5,2Dnum6,2Dnum7,serra,2Dnum8,2Dnum9,2Dnum10} Of
particular interest is the emergence of a Wigner
molecule,~\cite{wigmol1,wigmol2} the finite counterpart of a Wigner
crystal.~\cite{wigner} The high level of symmetry of typical 2D
quantum dots (such as circular dots or pillars) results in {\em
  rotating} Wigner molecules, with a rotationally invariant density
profile. As a result, the direct signatures of a Wigner molecule on
the electronic density of the system are
weak.~\cite{wigmol1,wigmol2,2Dnum8} Such states can be fully
characterized only either considering their roto-vibrational
spectrum~\cite{wigspec} or by considering density-density correlation
functions, whose experimental probe is particularly problematic. At
the experimental level Wigner molecules have thus essentially been
investigated by means of optical spectroscopic
techniques,~\cite{wigmol1,wigmol2} while attempts at imaging the
correlated electron wavefunction employing a scanning tunnel
microscope (STM) have been put forward.~\cite{maxwf}\\
\noindent Due to the reduced dimensionality, interaction effects in
one-dimensional (1D) quantum dots are even more dramatic. Indeed,
interaction and quantum confinement effects are directly visible at
the level of the electron density. Such 1D quantum dots can be
realized in several different ways, ranging from carbon
nanotubes~\cite{bockrath} (CNTs) to cleaved edge overgrowth (CEO)
wires.~\cite{amir} The quantum confinement within a region of length
$L$ causes Friedel oscillations,~\cite{vignale} with a typical
wavelength $\lambda_{F}=2\pi/k_{F}$ with $k_{F}$ the Fermi
momentum.~\cite{sablikov}\\
\noindent More intriguing is the formation of a Wigner molecule when
Coulomb interactions exceed the kinetic energy. In 1D it is {\em
  pinned} into the dot and induces a peculiar oscillatory
pattern~\cite{1Dwig} on the electron density with a wavelength
$\lambda_{W}=\lambda_{F}/2$.\\

\noindent One-dimensional quantum dots have been investigated by means
of numerical techniques, ranging from density functional approaches to
exact
diagonalizations,~\cite{kramer,pederiva,bedu,szafran,wire3,polini,shulenburger,secchi1,sgm2,astrak}
confirming the above picture. Despite their precision, such methods
suffer of some limitation. They are usually restricted to a low number
of particles and do not offer the flexibility of analytical models,
which allow to investigate issues such as transport properties more
easily.\\
\noindent An analytical approach, widely employed to describe the
low-energy sector of the physics of interacting 1D electrons is the
Luttinger liquid model.~\cite{giamarchi,voit} In connection with the
bosonization technique, it represents a powerful method to deal with
interacting 1D fermions, allowing to explore the limit of not too low
particle numbers. Interaction effects are modeled by non-universal
parameters $g_{\rho}$ and $g_{\sigma}$, the strength of interaction in
the charge and spin sectors respectively. This model has been applied
extensively to the study of the transport properties of 1D quantum
dots.~\cite{eggerimp,braggioepl,iotobias,kim,ioale1,ioale2,milena,graf}
It has also been applied to describe the formation of 1D Wigner
molecules.~\cite{k1/2,shulz,safi,sablikov,bortz}\\
\noindent Other models for 1D fermions map onto a Luttinger liquid in
their low-energy sector. The Hubbard model, for instance, can be
mapped onto a Luttinger liquid,~\cite{k1/2,bortz} with $g_{\rho}\geq
1/2$, while $g_{\sigma}=1$ due to SU(2) symmetry. The extended Hubbard
model removes the constraint on $g_{\rho}$.~\cite{k1/2} Models for
Wigner molecules consisting in anti-ferromagnetically coupled
electrons oscillating around their equilibrium positions
map~\cite{fiete1,glazman1,glazman2,075,fiete2} into a Luttinger liquid
with $g_{\sigma}=1$.\\
\noindent The Luttinger model suffers some drawbacks. One is connected
to the terms oscillating at wavelengths shorter than $\lambda_{F}$ in
the series of harmonics~\cite{haldane} of its electron density. In
general, the amplitudes of these terms are model-dependent and are
then free parameters.~\cite{safi} When including in the model terms
describing Wigner oscillations, indeed all the amplitudes become
weighted by phenomenological constants.~\cite{shulz} Only by comparing
with more refined methods one can attempt to determine such
constants.~\cite{bortz}\\
\noindent In addition, treating both charge and spin in the Luttinger
regime - the so called ``{\em spin-coherent}'' Luttinger model - is
strictly valid~\cite{fiete1,fiete2} only for temperatures and voltages
{\em smaller than the spin (and charge) bandwidth} $D_{\sigma}=N\pi
v_{\sigma}/L$ ($D_{\rho}=N\pi v_{\rho}/L$), where $v_{\sigma}$
($v_{\rho}$) is the velocity of spin (charge) excitations. When such a
constraint is not fulfilled, more refined models~\cite{075} should be
employed including the so called ``{\em spin-incoherent}''
liquid.~\cite{fiete1,fiete2}\\

\noindent Several methods are proposed and employed to experimentally
study Wigner molecules in 1D. Besides spectroscopical
tools,~\cite{transfer,chains} a Wigner molecule can be inferred from
the modifications induced on the transport
properties.~\cite{fili,nature} Since Wigner oscillations are present
in the electron density, it is also possible to directly probe such
quantities in, e.g., momentum-resolved tunneling experiments with
parallel quantum wires.~\cite{fili2,wire1,wire2,wire3}\\
\noindent {\em Local probes} are a promising technique. Recently, the
injection from a STM tip has been theoretically proposed to detect
local electron-vibron coupling,~\cite{noi}
Friedel~\cite{oscill,eggertstm,martin,bercprl,dolcini,nocera} or even
Wigner~\cite{secchi2} oscillations. An STM is however sensitive to the
tunneling density of states rather than to the electron density. More
suitable is a charged atomic force microscope (AFM) tip, already
proposed to image the spin-charge separation~\cite{lee,sgm1} in a
Luttinger liquid.~\cite{glazcap} Due to the capacitive coupling to the
electron density it allows to probe its oscillations. The effects of
an AFM tip on the energy levels of a 1D dot have been recently
considered theoretically.~\cite{sgm2,linear} However, the influence on
the conductance {\em amplitude} has not been addressed.\\

\noindent In this work we fill the above gap, studying a 1D quantum
dot described as a Luttinger liquid, capacitively coupled to an AFM
tip. The Luttinger model allows us to easily access the regime of
large particle numbers, not yet considered.~\cite{sgm2} We will
consider the ``spin-coherent'' regime. In full generality, we will
regard the interaction strength $g_{\rho}$ as a free parameter. The
Friedel and Wigner oscillations of the electron density will be fully
retained. We will develop a general and powerful framework which
allows to systematically investigate {\em both} the linear and
nonlinear transport properties to the lowest order in the tip
interaction strength, in the requested regime. In this work, we will
focus on the results concerning the {\em linear} transport regime. Our
main findings are the following.\\
\noindent Both the {\em position} and the {\em height} of the linear
conductance peak oscillate as a function of the tip position. While a
shift of the position of the linear conductance peak has been already
reported for small $N$,~\cite{sgm2,linear} the modulation of the {\em
  height} of the linear conductance peak is a novel result. These
oscillations bring information about the Friedel or Wigner
oscillations in the electron density. The oscillations induced by the
Wigner molecule act differently on the conductance peak position and
height as the interaction strength increases. In particular, while the
peak position oscillations due to Wigner get enhanced at strong
interactions, the peak height modulations are suppressed. Oscillations
due to Friedel, on the other hand, are robust against interaction both
in the position and in the amplitude modulations.\\

\noindent The scheme of the paper is the following. In
Sec.~\ref{sec:model} we outline the model describing the system in the
bosonized language. In Sec.~\ref{sec:chempot} we evaluate the
perturbative corrections to the dot chemical potential, while in
Sec.~\ref{sec:transport} we determine the tunneling rates in the
presence of the AFM tip and employ them to evaluate the transport
properties. Analytic expressions for the linear conductance are
provided in Sec.~\ref{sec:transport}. Section~\ref{sec:results}
contains our results . Finally, in Appendix~\ref{sec:appa} we outline
the evaluation of the tip-induced corrections to the tunneling rates.
\section{Model}
\label{sec:model}
\begin{figure}[htbp]
\begin{center}
\includegraphics[width=8cm,keepaspectratio]{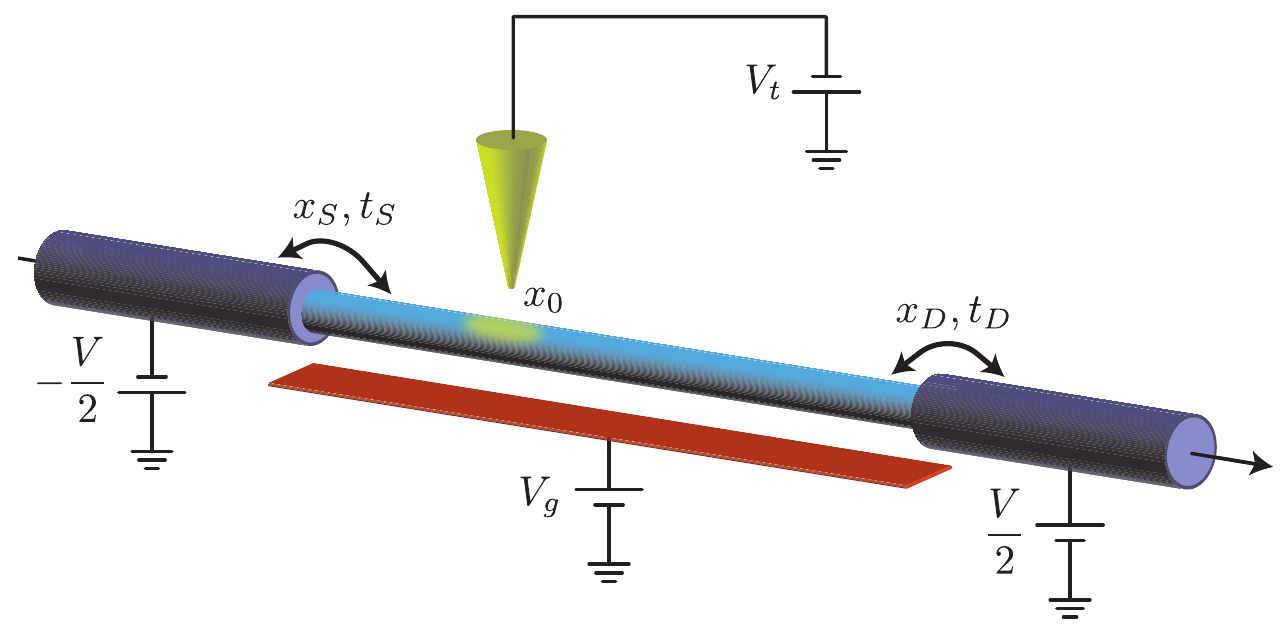}
\caption{(Color online) Schematic setup of the 1D quantum dot,
  perturbed by the negatively charged AFM tip ($V_{t}<0$) at position
  $x_0$, capacitively coupled to the gate contact potential $V_g$ and
  connected, via tunnel barriers at positions $x_{S}=0$, $X_{D}=L$
  with amplitudes $t_{S,D}$, to source $S$ and drain $D$ contacts,
  kept at a potential difference $V$.}
\label{fig:traverso1}
\end{center}
\end{figure}

The system under investigation - see Fig.~\ref{fig:traverso1} - is an
interacting one-dimensional (1D) quantum dot of length $L$
capacitively coupled to a negatively charged ($V_{t}<0$) atomic force
microscope (AFM) tip and tunnel-coupled to source ($S$) and drain
($D$) contacts. A gate contact, biased at $V_{g}$ and capacitively
coupled to the dot, is also included.\\
The system is described by the Luttinger model with a free band
linearized around the Fermi point $k_{F,s}=N_s^{(0)}\pi/L$, where
$N_{s}^{(0)}$ is the reference number of electrons with spin $z$
component $s=\pm$ in the ground state. We consider here a reference
state with an even total number $N_{+}^{(0)}=N_{-}^{(0)}$ and
introduce $k_{F}\equiv k_{F,s}$. The odd case can be reproduced adding
or subtracting one electron to the above case.

The dot Hamiltonian $H_{d}$ reads (from now on,
$\hbar=1$)
\begin{equation}
H_{d}=H_N+H_{b},\label{eq:toth}
\end{equation}
with~\cite{fabrizio}
\begin{eqnarray}
H_N&=&\frac{E_{\rho}}{2}( N_{\rho}-N_g)^2+\frac{E_{\sigma}}{2}N_{\sigma}^2, \label{eq:hn}\\
H_{b}&=&\sum_{n_{q}>0}\left[\varepsilon_{\rho}n_{q}d^\dag_{\rho,n_{q}}d_{\rho,n_{q}}+\varepsilon_{\sigma}n_{q}d^\dag_{\sigma,n_{q}}d_{\sigma,n_{q}}\right]\, . \label{eq:hp}
\end{eqnarray}
Here, $H_{N}$ represents the contribution of the zero modes
\begin{equation}
N_{\rho/\sigma}= N_{s=+}\pm  N_{s=-}\, ,
\end{equation}
with $N_{s}$ the number of extra electrons with spin $s$, with respect
to $N_{s}^{(0)}$. The energies $E_{\nu}=\pi v_{\nu}/2Lg_{\nu}$ have
been introduced in terms of the velocity $v_{\nu}$ of the mode
$\nu=\rho,\sigma$ and of the Luttinger parameters $g_{\nu}$. For
repulsive interactions one has $g_{\rho}=g<1$, while $g=1$ corresponds
to the noninteracting limit. On the other hand, $g_{\sigma}=1$ for an
SU(2) invariant theory.~\cite{voit} The velocity of the charged mode
is renormalized by the interactions and leads to $v_{\rho}=v_{F}/g$
where $v_{F}$ is the Fermi velocity. Within the standard Luttinger
theory, the spin velocity is $v_{\sigma}=v_{F}$ even if, in more
refined models, it may differ from this value.~\cite{fiete1,fiete2}
Furthermore, $N_{g}$ is the number of charges induced by a gate
electrode capacitively coupled to the dot.\\
\noindent Note that the parameter $E_{\rho}$ might deviate from the
simple expression quoted above due to e.g. screening induced by
surrounding gates. Therefore, $E_{\rho}$ will be in the following
treated as a free parameter with $E_{\rho}\gg E_{\sigma}$.\\
\noindent The term $H_{b}$ describes collective, quantized charge and
spin density waves with boson operators $d_{\nu,n_q}$ and
$\varepsilon_{\nu}=\pi v_{\nu}/L$, with $n_{q}\in\mathbb{N}$.\\
\noindent The electron field operator $\Psi_s(x)$ satisfying open
boundary conditions $\Psi_s(0)=\Psi_s(L)=0$ is
\begin{equation}
\Psi_s(x)=e^{ik_Fx}\psi_{s,+}(x)+e^{-ik_Fx}\psi_{s,-}(x)\, , \label{eq:optot}
\end{equation}
where $\psi_{s,r}(x)$ are $2L$-periodic fermion fields representing
right ($r=+$) and left ($r=-$) movers in the dot. Due to the open
boundaries conditions one has $\psi_{s,r}(x)=-\psi_{s,-r}(-x)$. The
right movers operator admits a bosonic representation~\cite{fabrizio}
\begin{equation}
\psi_{s,+}(x)=\frac{\eta_s}{\sqrt{2\pi\alpha}}e^{-i\theta_{s}}
\,e^{i\frac{\pi  N_sx}{L}}e^{i\frac{\Phi_{\rho}(x)+s\Phi_\sigma(x)}{\sqrt{2}}}\,. \label{eq:opright}
\end{equation}
Here, $\alpha$ is the cutoff length, of order $1/k_F$, $\theta_{s}$
satisfies $[\theta_{s}, N_{s'}]=i\delta_{s,s'}$, and $\eta_s$ fulfill
$ \eta_s\eta_{s'}+\eta_{s'}\eta_s=2\delta_{s,s'}$, allowing the right
anticommutation relations for different spins. The boson fields
$\Phi_{\rho}(x)$, $\Phi_{\sigma}(x)$ are given by
\begin{equation}
\Phi_{\nu}(x)=\sum_{n_{q}>0}\frac{e^{-\alpha q/2}}{\sqrt{g_{\nu}n_q}}
\left[\left(\cos{(qx)}-ig_{\nu}\sin{(qx)}\right)d^\dag_{\nu,n_{q}}+h.c.\right]\nonumber
\end{equation}
with $q=n_q\pi/L$.\\

The coupling with the AFM tip is modeled as a capacitive interaction
\begin{equation}
H_{tip}=-e\int_{0}^{L}dx\ V_{tip}(x)\rho(x)\label{eq:tipzero}
\end{equation}
between the electron density $-e\rho(x)$ and the tip potential
$V_{tip}(x)$, peaked around $x_0$.

\noindent The particle density $\rho(x)=\sum_{s=\pm}\rho_s(x)$ with
\begin{equation}
\rho_{s}(x)=\Psi_{s}^{\dagger}(x)\Psi_{s}(x)
\end{equation}
can be bosonized~\cite{comm} using
Eqns.~(\ref{eq:optot},\ref{eq:opright}), one has
($\alpha=k_F^{-1}$)

\begin{equation}
\rho_s(x)=\frac{k_F}{\pi}+\frac{N_{s}}{L}-\frac{1}{\pi}\partial_x \varphi_s(x)+\rho_s^F(x)\,. \label{eq:den2}
\end{equation}
Here, the first three terms represent the long wave part expressed in
terms of the antisymmetric field
\begin{eqnarray}\varphi_s(x)&=&\frac{\varphi_\rho(x)+s\varphi_\sigma(x)}{\sqrt{2}},\\
\varphi_{\rho/\sigma}(x)&=&\frac{1}{2}\left[
  \Phi_{\rho/\sigma}(-x)-\Phi_{\rho/\sigma}(x)\right]\,, \label{eq:varphi} \end{eqnarray}
while the Friedel part $\rho_s^F(x)$ is given by~\cite{shulz,fabrizio}
\begin{eqnarray}
\rho_s^F(x)&=&-\frac{k_F}{\pi}\cos\left[\mathcal{L}(N_{s},x)-2\varphi_{s}(x)\right]\, ,\\
\mathcal{L}(n,x)&=&2k_{F} x+\frac{2\pi x}{L}n -2h(x)\, .
\end{eqnarray}
with
\begin{equation}
h(x)=\frac{1}{2}\tan^{-1}\left[\frac{\sin(2\pi x/L)}{e^{\pi\alpha/L}-\cos (2\pi x/L)}\right]\, .
\end{equation}

In addition to the above terms, we will consider the so called Wigner
contribution
\begin{equation}
\rho^W(x)\propto e^{-4ik_Fx}\psi^\dag_{+,+}(x)\psi_{+,-}(x)\psi^\dag_{-,+}(x)\psi_{-,-}(x)+\mathrm{h.c.}\nonumber
\end{equation}
which arises from the presence of electron-electron
interaction~\cite{bortz} beyond the Luttinger approximation, external
perturbations,~\cite{safi} or by effects of ion-electron
interactions.~\cite{fiete1} Its bosonic form is~\cite{shulz}
\begin{equation}
\rho^W\!\!(x)\!=\!-\frac{A k_{F}}{\pi}\!\!\cos\!\left[2\mathcal{L}(N_{\rho}/2,x)-2\sqrt{2}\varphi_\rho(x)\right]\, ,
\end{equation}
where $A$ is a model dependent constant. Note that, in contrast to the
Friedel term, the Wigner one depends on the charge sector only. The
coupling with the tip is then represented as
$H_{tip}=H^{F}_{tip}+H^W_{tip}$, with
\begin{eqnarray}
H^F_{tip}&=&V_{F}\sum_s \cos\!\left[\mathcal{L}(N_{s},x_{0})-2\varphi_s(x_{0})\right]\label{eq:htipF}\\
H^W_{tip}&=&V_{W}\cos\!\left[2\mathcal{L}(N_{\rho}/2,x_{0})-2\sqrt{2}\varphi_\rho(x_{0})\right]\, , \label{eq:htipW}
\end{eqnarray}
where $V_{F}$ and $V_{W}$ are free parameters that depend on the shape of
the AFM tip and on the weight of the Friedel/Wigner oscillations. Note
that we neglected the coupling with the long wavelength part of the
density since it can be adsorbed in the Hamiltonian with a unitary
transformation.\\
\section{Chemical potential}
\label{sec:chempot}
The presence of $H_{tip}$ shifts the energy levels of the quantum dot
\begin{equation}
E(N_{\rho},N_{\sigma},x_{0})=E^0(N_{\rho},N_{\sigma})+\langle H_{tip}(x_{0})\rangle
\label{energy}
\end{equation}
with respect to the bare case
\begin{equation}
E^0({N_{\rho},N_{\sigma}})=\frac{E_{\rho}}{2}( N_{\rho}-N_g)^2+
\frac{E_{\sigma}}{2}N_{\sigma}^2\, ,
\end{equation}
with $\langle\ldots\rangle$ the thermal average with respect to
$H_{b}$ at fixed particle number. The average is decomposed in
Friedel and Wigner terms, for $T\to0$ one has
\begin{equation}
\langle H_{tip}(x_{0})\rangle=\sum_{\xi=F,W}E^{\xi}_{tip}(N_{\rho},N_{\sigma},x_{0})
\end{equation}
with
\begin{eqnarray}
\!\!\!\!\!\!E^{F}_{tip}(N_{\rho},N_{\sigma},x_{0})\!\!&=&\!\!V_{F}K^{F}(x_{0})\sum_{s=\pm}\cos\left[\mathcal{L}({N_s},x_{0})\right]\, ,\label{eq:enerF}\\
\!\!\!\!\!\!E^{W}_{tip}(N_{\rho},N_{\sigma},x_{0})\!\!&=&\!\!V_{W}K^W(x_{0})\cos\left[2\mathcal{L}({N_{\rho}/2},x_{0})\right]\, ,\label{eq:enerW}
\end{eqnarray}
where $K^{F}(x_{0})~=~\exp{[-2\langle\varphi_{s}^{2}(x_{0})\rangle]}$
and $K^{W}(x_{0})~=~\exp{[-4\langle\varphi_{\rho}^{2}(x_{0})\rangle]}$
with
\begin{eqnarray}
K^F(x_{0})&=&\left[\frac{\sinh\left({\frac{\pi\alpha}{2L}}\right)}{\sqrt{\sinh^2\left({\frac{\pi\alpha}{2L}}\right)+\sin^2\left(\frac{\pi x_0}{L}\right)}}\right]^{\frac{1+g}{2}}\, ,\label{eq:Kf} \\
K^W(x_{0})&=&\left[\frac{\sinh\left({\frac{\pi\alpha}{2L}}\right)}{\sqrt{\sinh^2\left({\frac{\pi\alpha}{2L}}\right)+\sin^2\left(\frac{\pi x_0}{L}\right)}}\right]^{2g}\label{eq:Kw}.
\end{eqnarray}

The chemical potential for a given configuration with $N_{\rho}$
charges and spin $N_{\sigma}$ is defined as
\begin{equation}
\mu_{d}(N_{\rho},N_{\sigma},x_{0})=E(N_{\rho}+1,N_{\sigma}\pm1,x_{0})-E(N_{\rho},N_{\sigma},x_{0})\label{eq:fullchempot}
\end{equation}
with $E(N)$ given in Eq.~(\ref{energy}). The above expression holds either for ground states, where $N_\rho=0$ ($N_\rho=1$) for even (odd) $N$ ($N$ being the total number of particles) and
\begin{equation}
|N_{\sigma}|=\frac{1-(-1)^{N_{\rho}}}{2}\, ,\label{eq:constr}
\end{equation}
and for excited states where $|N_{\sigma}|$ attains larger values with possibly $N_\rho>0$ ($N_\rho>1$) for an even (odd) $N$. The chemical potential can be decomposed as
\begin{equation}
\mu_{d}(N_{\rho},N_{\sigma},x_{0})=\mu_{0}(N_{\rho},N_{\sigma})+\delta\mu(N_{\rho},N_{\sigma},x_{0})\, ,\label{eq:decompo}
\end{equation}
where
\begin{equation}
\mu_{0}(N_{\rho},N_{\sigma})=E_{\rho}\left(\frac{1}{2}+N_{\rho}-N_{g}\right)+E_{\sigma}\frac{1\pm2 N_{\sigma}}{2}\label{eq:chempot0}
\end{equation}
is the bare dot chemical potential and
\begin{equation}
\delta\mu(N_{\rho},N_{\sigma},x_{0})=\sum_{\xi=F,W}\delta\mu^{\xi}(N_{\rho},N_{\sigma},x_0)\label{eq:deltamu}
\end{equation}
are the corrections due to the tip. They have been the subject of
numerical investigation~\cite{sgm2} with exact diagonalization
techniques in the regime of low $N_{\rho}$. As can be seen in
Eq.~(\ref{eq:enerF}) and Eq.~(\ref{eq:enerW}),
$\delta\mu^{\xi}(N_{\rho},N_{\sigma},x_{0})$ exhibits an oscillatory
shape enveloped by $K^{\xi}(x_{0})$.\\ Let us now specify the above
general expressions to the case involving ground states only,
i.e. when Eq.~(\ref{eq:constr}) holds, relevant to study the linear
transport regime. By exploiting the relation
$2N_{s}=N_{\rho}+sN_{\sigma}$ it is easy to show that
\begin{widetext}
\begin{eqnarray}
\delta\mu^{F}(x_{0})&=&V_{F}K^{F}(x_{0})\left\{\cos\left[\mathcal{L}(\mathcal{N}+1,x_{0})\right]-\cos\left[\mathcal{L}(\mathcal{N},x_{0})\right]\right\}\, ,\label{eq:deltamuF}\\
\delta\mu^{W}(x_{0})&=&V_{W}K^{W}(x_{0})\left\{\cos\left[2\mathcal{L}\left(\frac{N_{\rho}+1}{2},x_{0}\right)\right]-\cos\left[2\mathcal{L}\left(\frac{N_{\rho}}{2},x_{0}\right)\right]\right\}\, ,\label{eq:deltamuW}
\end{eqnarray}
\end{widetext}
with
\begin{equation}
\mathcal{N}=\begin{cases}
N_{\rho}/2 & \mathrm{if}\ N_{\rho}\ \mathrm{is\ even}\\
(N_{\rho}-1)/2 & \mathrm{if}\ N_{\rho}\ \mathrm{is\ odd}\\
\end{cases}\label{eq:ncal}
\end{equation}
The corrections to the chemical potential present an oscillatory
behavior given by the superposition of two cosine terms.
For the Friedel case, if $N$ is even they oscillate with wavelength $2L/(N+2)$ and $2L/N$, while for odd $N$ their wavelength is $2L/(N+1)$ and $2L/(N-1)$. For the Wigner case, one always finds the spacial frequencies $L/(N+1)$ and $L/N$ .\\
\noindent These oscillating patterns are modulated by the functions
$K^{F}(x_{0})$ and $K^{W}(x_{0})$. They depend both on the number of
electrons through the cut-off $\alpha=k_{F}^{-1}$ and on the
interaction parameter $g_{\rho}$, dictating their power-law scaling.\\
\noindent The interaction parameter $g$ dictates the relevance of the
Wigner term in comparison to the Friedel one. By inspecting
Eqns.~(\ref{eq:Kf},\ref{eq:Kw}) it is easy to show that Wigner
correlations become relevant for $g<1/3$, see Ref.~\onlinecite{safi}.

\noindent In addition, as $N$ grows, $\alpha$ and thus
$K^{\xi}(x_{0})$ are suppressed. In this limit quantum confinement
effects become less relevant as the system crosses over towards the
semi-classical regime. Furthermore, as the high-density limit is
approached kinetic terms become more relevant than Coulomb
repulsion,~\cite{vignale} leading to a further suppression of the
signatures due to the Wigner molecule.\\
\begin{figure}[htbp]
\begin{center}
\includegraphics[width=8cm,keepaspectratio]{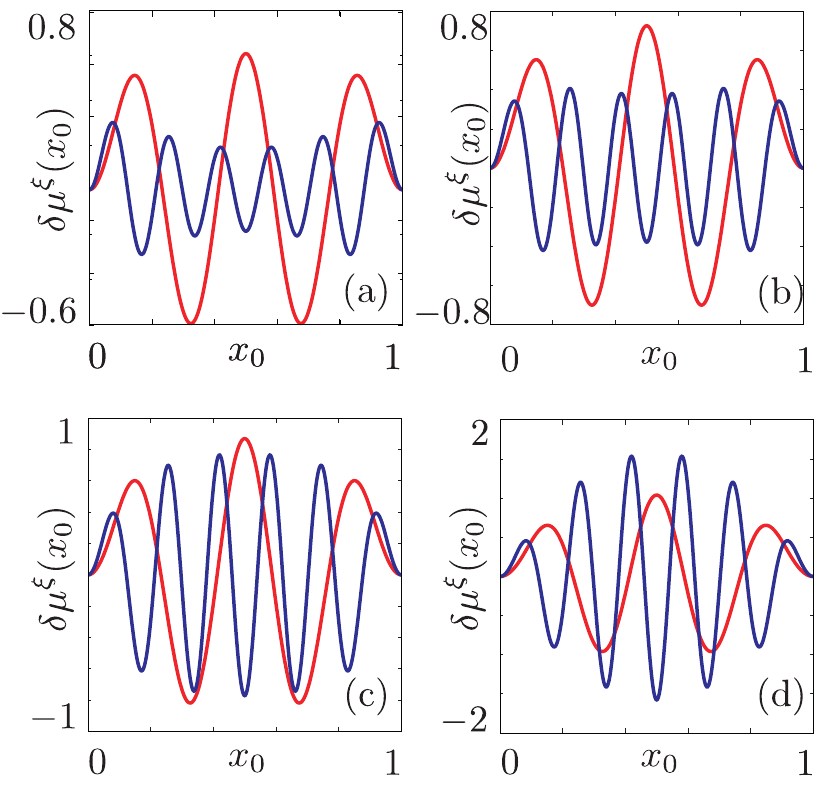}
\caption{(Color online) Chemical potential corrections
  $\delta\mu^{\xi}(x_{0})$ (units $E_{\rho}$) as a function of $x_{0}$
  (units $L$) for the transition $5\leftrightarrow 6$ and different
  values of the interaction parameter: (a) g=1; (b) g=0.7; (c) g=0.4;
  (d) g=0.1. Red (gray) curves represent the Friedel correction, blue
  (dark gray) curves the Wigner correction.  In all panels,
  $V_{F}=V_{W}=\varepsilon_{\sigma}$ and $\alpha=k_{F}^{-1}=L/3\pi$.}
\label{fig:traverso2}
\end{center}
\end{figure}
\begin{figure}[htbp]
\begin{center}
\includegraphics[width=8cm,keepaspectratio]{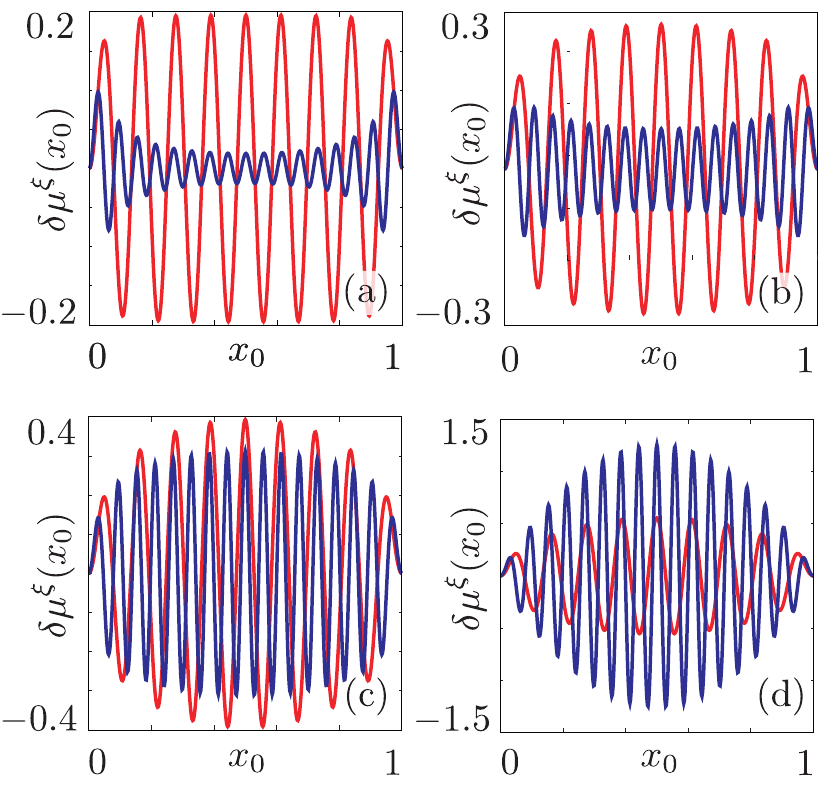}
\caption{(Color online) Same as in Fig.~\ref{fig:traverso2} but for
  $16\leftrightarrow 17$ and $\alpha=k_{F}^{-1}=\pi/8L$.}
\label{fig:traverso3}
\end{center}
\end{figure}
The above analysis is confirmed by Figs.~\ref{fig:traverso2}
and~\ref{fig:traverso3} which show the corrections to the chemical
potential for $N=5$ and $N_{\rho}=16$ respectively, and
different values of $g$. Indeed, for each of these cases it is clear
that Wigner oscillations grow and eventually become relevant as $g\to
0$. Also, it is clear that for a given interaction strength both
Friedel and Wigner oscillations get smaller as the number of partiles
increases.
\section{Transport}
\label{sec:transport}
The coupling between the dot and the leads is produced by tunneling
barriers at $x_S=0$ and $x_D=L$
\begin{equation}
H_t=\sum_{\lambda=S,D}\sum_{s=\pm}\left[t_{\lambda}\chi^\dag_{s,\lambda}(x_{\lambda})\psi_{s,+}(x_{\lambda})+h.c.\right]=\sum_{\lambda}H_{t}^{\lambda}\, ,\nonumber
\end{equation}
with $\chi_{s,\lambda}(x_{\lambda})$ ($\lambda=S,D$) the fermion
operators at the end point of the lead $\lambda$ and $t_{\lambda}$ the
transmission amplitudes of the tunnel barriers. The dot is subject to
a symmetric voltage drop $V_{S}=-V/2$, $V_{D}=V/2$ between the two
leads. The leads are modeled as non-interacting Fermi gases with
$H_{leads}=H_{S}+H_{D}$ and we have set the bare leads chemical
potential to zero.\\

The key quantities of transport in the sequential regime are the
tunneling rates. They will be evaluated via the time evolution of the
density operator.\\

\noindent At the initial time $t=0$ the dot is characterized by the
state $|{\mathcal
  S}^{i}\rangle\equiv|N_{\rho}^{i},N_{\sigma}^{i}\rangle$. Here, we
will consider the full dynamics of the zero modes while the bosonic
excitations will be assumed in thermal equilibrium. The initial
density operator then reads
\begin{equation}
\rho(0)=\frac{e^{-\beta \left(H_S+H_D\right)}}{Z_SZ_D}\frac{e^{-\beta H_b}}{Z_{b}}|\mathcal{S}^{i}\rangle\langle\mathcal{S}^{i}|\, ,
\end{equation}
with the partition functions $Z_r=Tr\left\{e^{-\beta H_r}\right\}$,
$r\in\{b,S,D\}$ and $\beta=1/(k_BT)$. The probability ${\mathcal
  P}_{{{\mathcal S}^i} \rightarrow {{\mathcal{S}}^f}}(t)$ of finding
the dot in the final state $|\mathcal{S}^f\rangle=|N^f_\rho,
N^f_\sigma\rangle$ at time $t$ is then obtained tracing out the leads
and the bosonic degrees of freedom
\begin{equation}
\mathcal{P}_{{{\mathcal{S}}^i}\rightarrow { {\mathcal{S}}^f}}(t)=Tr\langle  {\mathcal{S}}^f|\rho(t)|{\mathcal{S}}^f\rangle\, ,\label{eq:prob}
\end{equation}
with $\rho(t)$ in the interaction picture~\cite{blum} with respect to
$H^{(0)}=H_{d}+H_{leads}$
\begin{equation}
\rho(t)=\left[Te^{-i\int_0^\tau d\tau {\cal V}(\tau)}\right]\rho(0)\left[\widetilde Te^{i\int_0^\tau d\tau {\cal V}(\tau)}\right]\, .
\end{equation}
Here, ${\mathcal
  V}(\tau)=e^{iH^{(0)}\tau}(H_{tip}+H_t)e^{-iH^{(0)}\tau}$ and $T$
($\widetilde T$) the time (anti-time) ordering operators.  The
probability in Eq.~(\ref{eq:prob}) is computed to the lowest order in
the tunneling barriers and in the coupling to the AFM tip. The typical
structure is
\begin{equation}
\mathcal{P}_{{{\mathcal{S}}^i}\rightarrow { {\mathcal{S}}^f}}(t)=\sum_{\lambda=S,D}\mathcal{P}_{{{\mathcal{S}}^i}\rightarrow { {\mathcal{S}}^f}}^{\lambda}(t)
\end{equation}
with $\mathcal{P}_{{{\mathcal{S}}^i}\rightarrow { {\mathcal{S}}^f}}^{\lambda}(t)=\mathcal{P}_{{{\mathcal{S}}^i}\rightarrow { {\mathcal{S}}^f}}^{0,\lambda}(t)+\delta\mathcal{P}_{{{\mathcal{S}}^i}\rightarrow { {\mathcal{S}}^f}}^{\lambda}(t)$ and
\begin{widetext}
\begin{eqnarray}
\mathcal{P}_{{{\mathcal{S}}^i}\rightarrow { {\mathcal{S}}^f}}^{0,\lambda}(t)&\!=\!&\int_0^tdt_1\int_0^tdt_2F_0^{\lambda}(t_1,t_2),\label{eq:PP0}\\
\delta\mathcal{P}_{{{\mathcal{S}}^i}\rightarrow { {\mathcal{S}}^f}}^{\lambda}(t)&\!=\!&\int_0^t\!dt_1\!\int_0^t\!dt_2\!\int_0^{t_2}\!\!\!dt_3\!\!\! \sum_{i=1,2}\!F_i^{\lambda}(t_1,t_2,t_3)+c.c.\ ,\label{eq:PP1}
\end{eqnarray}
\end{widetext}
where
\begin{eqnarray}
\hspace{-1.0cm}F_0^{\lambda}(t_1,t_2)&\!\!\!=\!&Tr\langle  \mathcal{S}^f|H_{t}^{\lambda}(t_1)\rho(0)H_{t}^{\lambda}(t_2)|\mathcal{S}^f\rangle,\label{eq:F0}\\
\hspace{-0.7cm}F_1^{\lambda}(t_1,\!t_2,\!t_3)&\!\!\!=\!\!&iTr\langle \mathcal{S}^f|H_t^{\lambda}(t_1)\rho(0)H_{t}^{\lambda}(t_3)H_{tip}(t_2)|\mathcal{S}^f\rangle,\label{eq:F1}\\
\hspace{-0.7cm}F_2^{\lambda}(t_1,\!t_2,\!t_3)&\!\!\!=\!\!&iTr\langle \mathcal{S}^f|H_t^{\lambda}(t_1)\rho(0)H_{tip}(t_3)H_{t}^{\lambda}(t_2)|\mathcal{S}^f\rangle\, .\label{eq:F2}
\end{eqnarray}
Note that in the sequential tunneling limit considered in this paper
the selection rules
$|N_{\rho}^{f}-N_{\rho}^{i}|=|N_{\sigma}^{f}-N_{\sigma}^{i}|=1$
apply. In the following, we will focus on the case where only two
charge states are involved in the transport.\\
The rates are given by the long time limit of the time derivate of
$\mathcal{P}^{\lambda}(t)$
\begin{equation}
\Gamma_{\mathcal{S}^{i}\to\mathcal{S}^{f}}^{\lambda}=\lim_{t\rightarrow\infty}\dot{\mathcal{P}}_{\mathcal{S}^{i}\to\mathcal{S}^{f}}^{\lambda}(t)\, .\label{eq:pdotrate}
\end{equation}
Their explicit evaluation, within the bosonization formalism, is presented in detail in Appendix~\ref{sec:appa}. Here we quote the main results. The rates have the general expression
\begin{equation}
\Gamma_{\mathcal{S}^{i}\to\mathcal{S}^{f}}^{\lambda}=\Gamma_{\mathcal{S}^{i}\to\mathcal{S}^{f}}^{0,\lambda}+\delta\Gamma_{\mathcal{S}^{i}\to\mathcal{S}^{f}}^{F,\lambda}(x_{0})+\delta\Gamma_{\mathcal{S}^{i}\to\mathcal{S}^{f}}^{W,\lambda}(x_{0})\, ,\label{eq:rategen}
\end{equation}
with
\begin{equation}
\Gamma_{\mathcal{S}^{i}\to\mathcal{S}^{f}}^{0,\lambda}=\Gamma_{0}^{\lambda}\sum_{q_{\rho},q_{\sigma}}A_{q_{\rho},q_{\sigma}}f\left[\Delta E_{\lambda}+q_{\rho}\varepsilon_{\rho}+q_{\sigma}\varepsilon_{\sigma}\right]
\end{equation}
representing the tunneling rate in the absence of the tip apart from
the {\em full} chemical potential including tip corrections and
\begin{equation}
\delta\Gamma_{\mathcal{S}^{i}\to\mathcal{S}^{f}}^{\xi,\lambda}(x_{0})=2V_{\xi}\Gamma_{0}^{\lambda}K^{\xi}(x_{0})\sum_{q_{\rho},q_{\sigma}}A_{q_{\rho},q_{\sigma}}R_{q_{\rho},q_{\sigma}}^{\xi,\lambda}(x_{0})
\end{equation}
being the {\em explicit} corrections induced by the tip. Here,
$\Gamma_{0}^{\lambda}=\nu_{0}\left|t_{\lambda}\right|^2/\pi\alpha$
with $\nu_{0}$ the leads density of states and $\Delta
E_{\lambda}=\Delta n
eV_{\lambda}+\mu_{d}(N_{\rho}^{i},N_{\sigma}^{i},x_{0})$ with $\Delta
n=(N_{\rho}^{f}-N_{\rho}^{i})$ and
$\mu_{d}(N_{\rho}^{i},N_{\sigma}^{i},x_{0})$ defined in
Eq.~(\ref{eq:fullchempot}). The coefficients $A_{q_{\rho},q_{\sigma}}$
and $R_{q_{\rho},q_{\sigma}}^{\xi,\lambda}(x_{0})$ are respectively
defined in Eq.~(\ref{eq:aqrho}) and Eq.~(\ref{eq:TheRate}).\\
\noindent The above rates fulfill the detailed balance relation
\begin{equation}
\Gamma_{\mathcal{S}^{i}\to\mathcal{S}^{f}}^{\lambda}=e^{-\beta\Delta E_{\lambda}}\Gamma_{\mathcal{S}^{f}\to\mathcal{S}^{i}}^{\lambda}\, .\label{eq:detbal}
\end{equation}
\noindent The framework developed here is very general and allows to
address both linear and nonlinear transport.~\cite{haupt,piovano} In
the rest of the paper, however, we will focus on the {\em linear}
transport regime which already shows a rich and interesting
physics. The nonlinear regime will be the subject of forthcoming
investigations.\\

\noindent In the linear regime only two charge values $N$ and $N+1$ will be considered, with the correponding ground state spins. Namely,
\begin{eqnarray}
|N_{\rho}=0,N_{\sigma}=0\rangle\quad&;&\quad|N_{\rho}=1,N_{\sigma}=\pm 1\rangle\ (N\ \textrm{even})\, ,\nonumber\\
|N_{\rho}=0,N_{\sigma}=\pm 1\rangle\quad&;&\quad|N_{\rho}=1,N_{\sigma}=0\rangle\ (N\ \textrm{odd})\, .\nonumber\\
\end{eqnarray}
Note that in the absence of magnetic field, states with
$N_{\sigma}=\pm 1$ are degenerate.\\
\noindent The standard expression for the linear conductance,
expressed in terms of the rates involving the above states,
is~\cite{braggioepl}
\begin{equation}
G=\left.\frac{\beta e^2D_{N}D_{N+1}\Gamma_{N\to N+1}^{S}\Gamma_{N+1\to N}^{D}}{D_{N}\Gamma_{N+1\to N_{\rho}}+D_{N+1}\Gamma_{N\to N+1}}\right|_{V=0}\, ,\label{eq:thelincond}
\end{equation}
where $\Gamma_{N_{\rho}\to N_{\rho}'}^{\lambda}$ is a shorthand
notation for
$\Gamma_{\mathcal{S}^{i}\to\mathcal{S}^{f}}^{\lambda}$,
$D_{N}=\left[3+(-1)^{N+1}\right]/2$ is the degeneracy of
the dot ground state with $N$ electrons and
\begin{equation}
\Gamma_{N_{\rho}\to N_{\rho}'}=\Gamma_{N_{\rho}\to  N_{\rho}'}^{S}+\Gamma_{N_{\rho}\to N_{\rho}'}^{D}\, .\nonumber
\end{equation}
\noindent The behavior of the conductance as a function of the
external parameters will be numerically investigated in detail in the
next section. Here we discuss a useful analytical approximation valid
at low temperatures around the maximum of the conductance peak which
is centered~\cite{beenakker} at $\mu_{d}(N_{\rho},N_{\sigma},x_{0})=0$
for $T=0$. As shown in Appendix~\ref{sec:appa}, in this regime the
tunneling rates can be approximated as
\begin{equation}
\Gamma_{N_{\rho}\to N_{\rho}+1}^{\lambda}=\gamma^{\lambda}(x_{0})f(\mu)\ ;\ \Gamma_{N_{\rho}+1\to N_{\rho}}^{\lambda}=\gamma^{\lambda}(x_{0})f(-\mu)\nonumber\\
\end{equation}
where $f(E)$ is the Fermi function,
$\mu\equiv\mu_{d}(N_{\rho},N_{\sigma},x_{0})$ is the dot chemical
potential and $\gamma^{\lambda}(x_{0})$ is defined in
Eq.~(\ref{eq:usefulcor}). Employing the above expressions and the
detailed balance in Eq.~(\ref{eq:detbal}), one can rewrite the
conductance as
\begin{equation}
G\approx\beta e^2D_{N}\Delta(x_{0})\frac{f(-\mu)}{1+(D_{N}/D_{N+1})e^{\beta\mu}}\, ,\label{eq:finalpeak}
\end{equation}
with
\begin{equation}
\Delta(x_{0})=\frac{\gamma^{S}(x_{0})\gamma^{D}(x_{0})}{\gamma^{S}(x_{0})+\gamma^{D}(x_0)}\, .
\end{equation}
Equation~(\ref{eq:finalpeak}) describes a resonance peak located at
\begin{equation}
\mu=\frac{k_{B}T}{2}\log\left(\frac{D_{N+1}}{D_{N}}\right)\, .
\end{equation}
When expressed in terms of $N_{g}$, the resonance is at
\begin{equation}
N_{g}=N_{g}^{*}+\frac{\delta\mu(x_{0})}{E}\label{eq:muosc}
\end{equation}
with
\begin{equation}
N_{g}^{*}=N+\frac{1}{2}+\frac{E_{\sigma}}{2E_{\rho}}(-1)^{N_{\rho}}-\frac{k_{B}T}{2}\log{\left(\frac{D_{N_{\rho}+1}}{D_{N_{\rho}}}\right)}
\end{equation}
and $\delta\mu(x_{0})$ given in Eq.~(\ref{eq:deltamu}). Therefore, the
position of the conductance oscillates due to the tip-induced spatial
fluctuations of the chemical potential.\\
\noindent Besides the oscillations of the peak position, also the {\em
  amplitude} of the conductance peak exhibits a modulation depending
on the location of the tip: on resonance, the conductance evaluates to
\begin{equation}
G_{\mathrm{res}}^{}=\beta e^{2}\frac{D_{N}D_{N+1}}{\left(\sqrt{D_{N}}+\sqrt{D_{N+1}}\right)^2}\Delta(x_{0})\, ,
\end{equation}
and thus explicitly depends on $x_{0}$ via the term
$\Delta(x_{0})$. Such modulation has not been reported so
far.~\cite{sgm2,linear} The term $\Delta(x_{0})$ can be expanded to
within linear terms in $\delta\Gamma^{\xi,\lambda}$. Assuming
symmetric tunnel barriers with $\Gamma_{0}^{\lambda}=\Gamma_{0}$, one
has
$\Delta(x_{0})\approx\Delta_{0}+\Delta_{1}^{F}(x_{0})+\Delta_{1}^{W}(x_{0})$
with
\begin{equation}
\Delta_{0}=\frac{\Gamma_{0}}{2}\left(1-e^{\frac{-\pi\alpha}{L}}\right)^{-\frac{1+g}{2}}\, ,
\end{equation}
and
\begin{equation}
\Delta_{1}^{\xi}(x_{0})=2\Delta_{0}V_{\xi}K^{\xi}(x_{0})\sum_{\nu=c,s}F_{\nu}^{\xi}(x_{0})g_{\nu}^{\xi}(x_{0})\, .\label{eq:delta1}
\end{equation}
Here,
\begin{widetext}
\begin{eqnarray}
\!\!\!\!\!\!g_{c}^{F}(x_{0})&=&\cos{\left[\frac{2\pi(\mathcal{N}+1)}{L}x-2h(x)+2k_{F}x\right]}+\cos{\left[\frac{2\pi\mathcal{N}}{L}x-2h(x)\right]}\, ,\label{eq:gc}\\
\!\!\!\!\!\!g_{s}^{F}(x_{0})&=&\sin{\left[\frac{2\pi(\mathcal{N}+1)}{L}x-2h(x)+2k_{F}x\right]}-\sin{\left[\frac{2\pi\mathcal{N}}{L}x-2h(x)\right]}\, ,\label{eq:gs}
\end{eqnarray}
\end{widetext}
with $\mathcal{N}$ defined in Eq.~(\ref{eq:ncal}). The functions
$g_{c,s}^{W}$ have the same functional form after the replacements
$\mathcal{N}\to N$ and $h(x)\to2h(x)$. The weighting functions
$F_{c,s}(x_{0})$ are given by
\begin{eqnarray}
F_{c}^{\xi}(x_{0})&=&\sum_{\mathbf{m}}\frac{B_{\mathbf{0},\mathbf{m}}^{\xi}}{\Lambda}\left\{\cos(kx_{0})+\cos\left[k(1-x_{0})\right]\right\}\, ,\label{eq:Fc}\\
F_{s}^{\xi}(x_{0})&=&\sum_{\mathbf{m}}\frac{B_{\mathbf{0},\mathbf{m}}^{\xi}}{\Lambda}\left\{\sin(kx_{0})-\sin\left[k(1-x_{0})\right]\right\}\, ,\label{eq:Fs}
\end{eqnarray}
with $\mathbf{m}=\{m_{1},m_{2},m_{3},m_{4}\}$, $m_{i}\in\mathbb{N}$
and $m_{i}\geq 0$, $B_{\mathbf{n},\mathbf{m}}^{\xi}$ defined in
Eqns.~(\ref{eq:Bf},\ref{eq:Bw}),
$\Lambda=\varepsilon_{\rho}(m_{1}+m_{2})+\varepsilon_{\sigma}(m_{3}+m_{4})$,
$k=\pi(m_{1}-m_{2}+m_{3}-m_{4})/L$ and the summations are extended
over the set of $\mathbf{m}\neq\{0,0,0,0\}$.\\
\noindent It can be shown that $\Delta_{1}^{\xi}(x_{0})=\Delta_{1}^{\xi}(L-x_{0})$
and that $\Delta_{1}^{\xi}(L/2)=0$, therefore the amplitude modulations are
symmetric around the center of the dot and vanish there.
\section{Results}
\label{sec:results}
In this section we discuss the behavior of the linear conductance in
Eq.~(\ref{eq:thelincond}). The full numerical results will be
interpreted with the aid of the analyltical expressions developed in
Sec.~\ref{sec:transport}.\\
\noindent We start considering a weakly interacting dot.
\begin{figure}[htbp]
\begin{center}
\includegraphics[width=8cm,keepaspectratio]{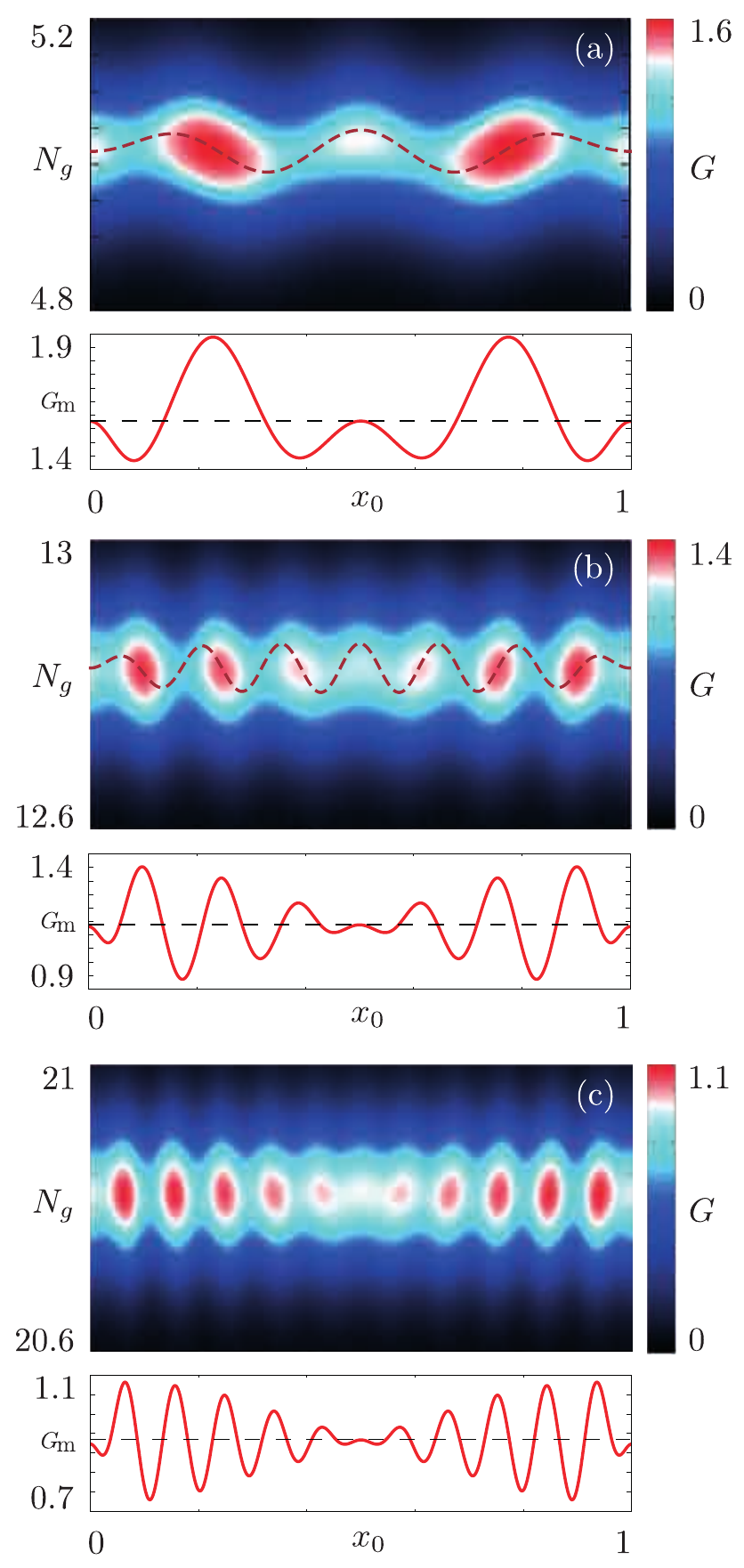}
\caption{(Color online) Color-map plot of the linear conductance $G$
  (units $e^2\Gamma_{0}/\varepsilon_{\sigma}$) as a function of
  $N_{g}$ and $x_{0}$ (units $L$) and plot of the conductance peak
  maximum $G_{\mathrm{m}}$ as a function of $x_{0}$ (units $L$) for a
  transition $N\leftrightarrow N+1$ with (a) $N=5$; (b) $N=12$; (c)
  $N=20$. The dashed lines superimposed to the color-maps represent
  the calculated chemical potential traces - see
  Eqns.~(\ref{eq:deltamu},\ref{eq:deltamuF},\ref{eq:deltamuW}) - the
  ones in the amplitude plots signal the value of the bare conductance
  peak obtained in the absence of the tip. Here, only the Friedel
  corrections have been included, with
  $V_{F}=0.07\varepsilon_{\sigma}$ and $V_{W}=0$. Other parameters are
  $k_{B}T=0.05\varepsilon_{\sigma}$, $E_{\rho}=5E_{\sigma}$ and
  $g=0.9$.}
\label{fig:traverso4}
\end{center}
\end{figure}
Figure~\ref{fig:traverso4} shows the conductance for $g=0.9$ and three
transitions $5\leftrightarrow 6$, $12\leftrightarrow 13$ and
$20\leftrightarrow 21$. For this calculation we have included {\em
  only} the corrections due to Friedel oscillations as one can expect
that the Wigner term represents a vanishing perturbation due to
$V_{W}\to 0$ in this limit.~\cite{bortz} The main panels show the
conductance with a colorscale plot, as a function of the tip position
$x_{0}$ and $N_{g}$. Below the density plot the maximum of the linear
conductance, $G_{\mathrm{m}}$, is shown as a function of $x_{0}$.  Two
main features are observed, namely\\

\noindent ($i$) oscillations of its position,\\
\noindent ($ii$) oscillations of its height.\\

\noindent The fact ($i$) is in agreement with previous
findings.~\cite{sgm2,linear} The fluctuations of the conductance peak
position turn out to be proportional to the tip-induced correction to
the chemical potential $\delta\mu(x_{0})\equiv\delta\mu^{F}(x_{0})$ in
this case - see Eqns.~(\ref{eq:deltamu},\ref{eq:deltamuF}). Indeed,
the oscillations of the resonance position exhibit a number of maxima and minima in accordance with the discussion in Sec.~\ref{sec:chempot}. This confirms the analytical
prediction shown in Sec.~\ref{sec:transport}, see
Eq.~(\ref{eq:muosc}).  As a specific example, for the case of the
transition $5\leftrightarrow 6$ the
plot in panel (a) exhibits three maxima and two minima.\\

\noindent More interesting is the {\em modulation of the height} of
the linear conductance which has never been reported so far. This
modulation is sizeable and could be detected in a transport
experiment. It exhibits a symmetric, {\em beating-like pattern}. The beating
phenomenon is particularly evident as $N$ increases. This fact will be
analyzed in more detail later in this section.
\begin{figure}[htbp]
\begin{center}
\includegraphics[width=8cm,keepaspectratio]{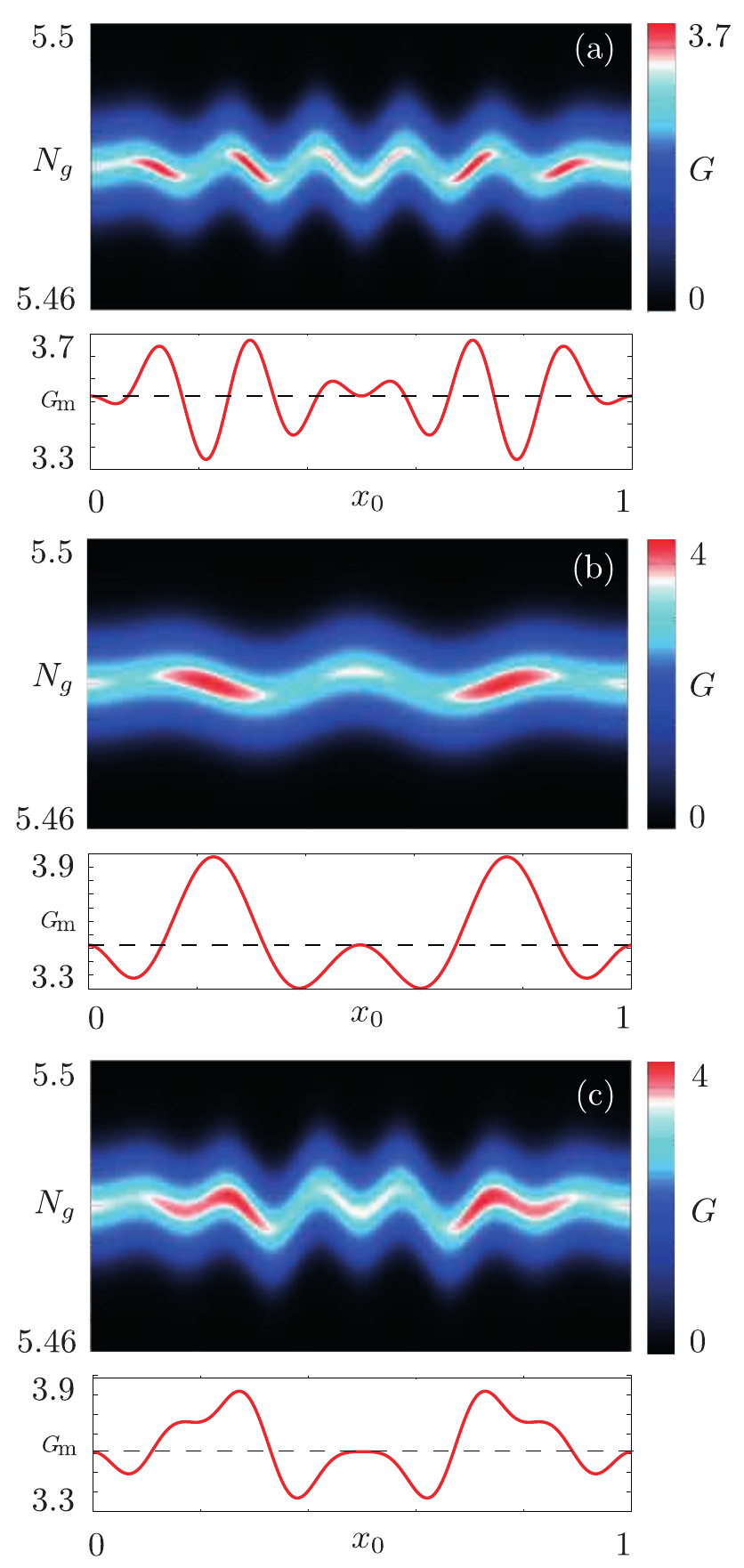}
\caption{(Color online) Color-map plot of the linear conductance
  (units $e^2\Gamma_{0}/\varepsilon_{\sigma}$) for the transition
  $N=5\leftrightarrow 6$ as a function of $N_{g}$ and $x_{0}$ (units
  $L$) and plot of the conductance peak maximum $G_{\mathrm{m}}$ as a
  function of $x_{0}$ (units $L$) calculated with (a) $V_{F}=0$,
  $V_{W}=0.07\varepsilon_{\sigma}$; (b)
  $V_{F}=0.07\varepsilon_{\sigma}$, $V_{W}=0$; (c)
  $V_{F}=V_{W}=0.07\varepsilon_{\sigma}$. Other parameters are
  $k_{B}T=0.05\varepsilon_{\sigma}$, $E_{\rho}=5E_{\sigma}$ and
  $g=0.2$.}
\label{fig:traverso5}
\end{center}
\end{figure}
\begin{figure}[htbp]
\begin{center}
\includegraphics[width=8cm,keepaspectratio]{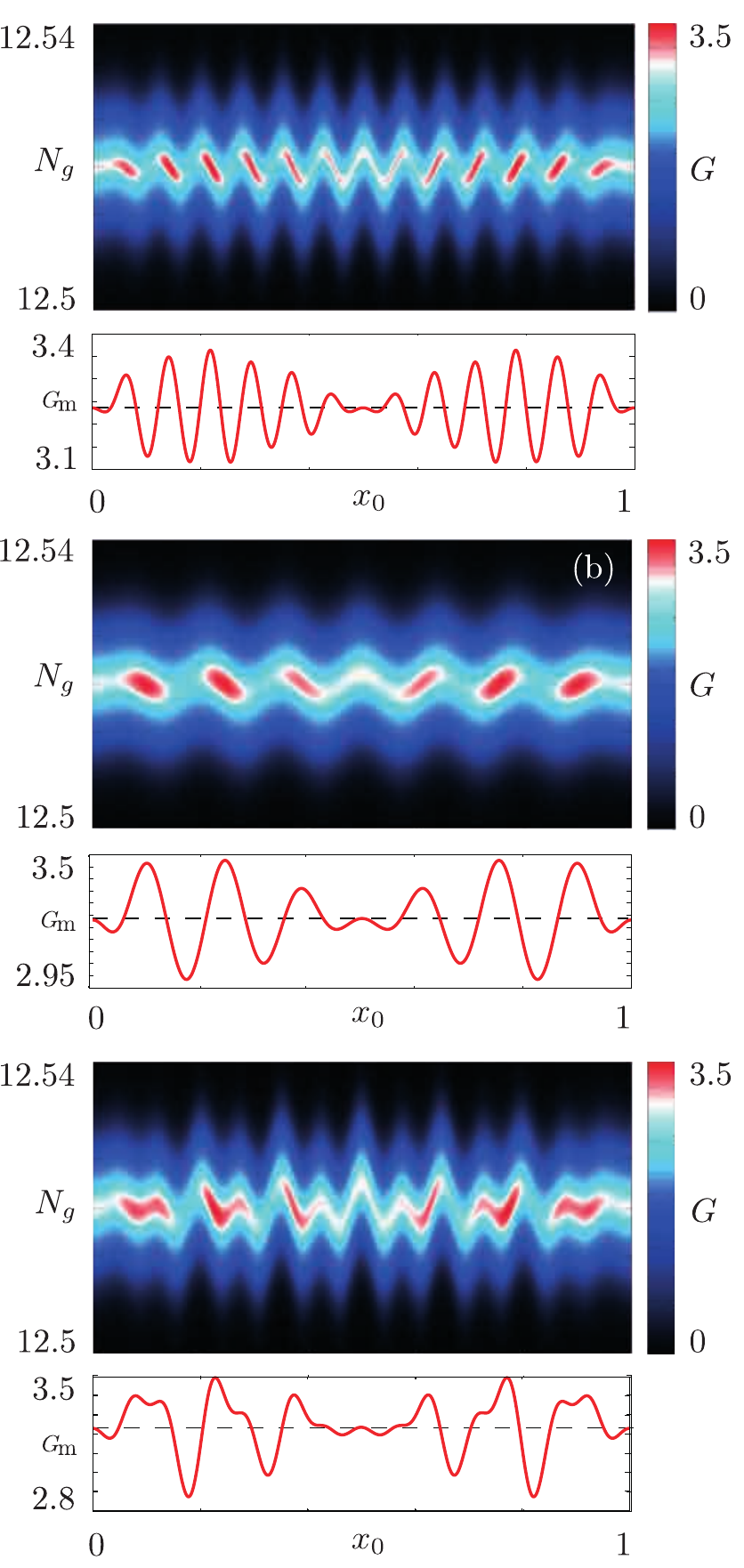}
\caption{(Color online) Same plot as in Fig.~\ref{fig:traverso5} but for
  the transition $N=12\leftrightarrow 13$ for (a) $V_{F}=0$,
  $V_{W}=0.07$; (b) $V_{F}=0.07$, $V_{W}=0$; (c)
  $V_{F}=V_{W}=0.07\varepsilon_{\sigma}$. Other parameters are
  $k_{B}T=0.05\varepsilon_{\sigma}$, $E_{\rho}=5E_{\sigma}$ and
  $g=0.2$.}
\label{fig:traverso6}
\end{center}
\end{figure}

\noindent Let us now turn to the case of stronger interactions and
study the effects due to the Wigner
corrections. Figures~\ref{fig:traverso5} and \ref{fig:traverso6} show the linear
conductance for $g=0.3$ and the transitions $5\leftrightarrow 6$ and
$12\leftrightarrow 13$ respectively. Panels (a) and (b) show the
Wigner and the Friedel contributions to the conductance. While the
Friedel corrections follow the pattern discussed above, Wigner
corrections exhibit a shorter wavelength with precisely $N+1$ maxima
and $N$ minima, double the number of those observed in the Friedel
case. The position of the conductance peak in the Wigner case follows
the position-dependent correction to the chemical potential
$\delta\mu^{W}(x_{0})$ defined in Eq.~(\ref{eq:deltamuW}). Friedel and
Wigner corrections induce therefore oscillatory patterns with very
different wavelengths, allowing to distinguish the two mechanisms in a
non-ambiguous way. Although we have shown the two contributions
separately, it can be in general expected that the Wigner and the
Friedel mechanisms co-exist. Therefore, in panels (c) we show the
conductance in the presence of both perturbations, considering the
case of equal amplitudes. Even if here we have chosen the same weights
for the Friedel and the Wigner terms ($V_{F}=V_{W}$), the latter
impresses its peculiar oscillatory pattern with wavelength $\approx
L/N$ on the conductance peak position. Indeed, as we have already
observed in Sec.~\ref{sec:chempot}, when interactions increase the
Wigner contribution to the chemical potential becomes even more
relevant than the Friedel one. This explains the persistence of $N+1$
maxima and $N$ minima in the oscillations of the chemical
potential. Also conductance amplitudes exhibit a pattern reminiscent
of the $N+1$ and $N$ maxima and minima but less pronounced with
respect to the one shown in the position of the conductance peak.\\

\noindent Let us now investigate the stability of these findings as a
function of the strength of interactions.
\begin{figure}[htbp]
\begin{center}
\includegraphics[width=8cm,keepaspectratio]{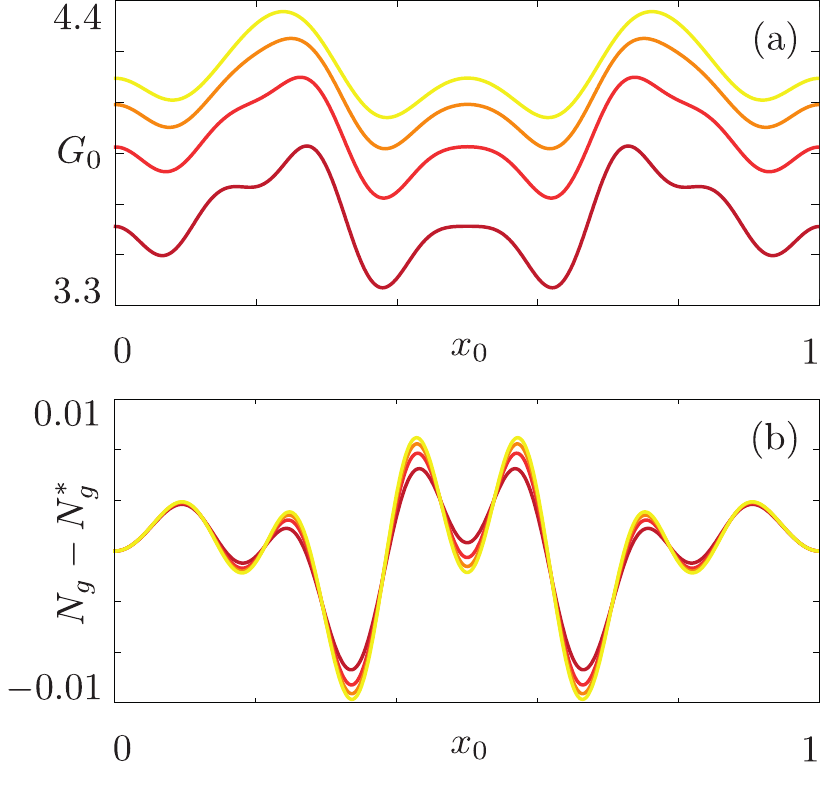}
\caption{(Color online) For the transition $5\leftrightarrow 6$ (a)
  Conductance maximum $G_{\mathrm{m}}$ (units
  $e^2\Gamma_{0}/\varepsilon_{\sigma}$) and (b) position offset
  $N_{g}-N_{g}^{*}$ - see Eq.~(\ref{eq:muosc}) - of the conductance
  peak as a function of $x_{0}$ (units $L$) and several values of the
  interaction parameter $g=0.2,\ 0.1,\ 0.05,\ 0.02$ (from darker to
  lighter color). Here we have chosen
  $V_{F}=V_{W}=0.07\varepsilon_{\sigma}$. Other parameters as in
  Fig.~\ref{fig:traverso5}.}
\label{fig:traverso7}
\end{center}
\end{figure}
\noindent Figure~\ref{fig:traverso7} shows the linear conductance
maximum and position for the transition $5\leftrightarrow 6$ as a
funciton of $x_{0}$ for several values of the Luttinger parameter
$g$. As interactions grow, and $g\to 0$, two striking features are
observed. While the oscillations of the chemical potential always
exhibit six maxima and five minima, with an increase in the
peak-to-valley ratio, the oscillations due to the {\em Wigner}
contribution in the peak {\em amplitude} tend to vanish.\\
\noindent Therefore, the effect of the Wigner corrections to the
chemical potential and to the peak amplitude are very different: in
the former case Wigner contributions are stable against Coulomb
interactions, while Wigner corrections to the conductance maximum
become vanishing for $g\to 0$.

\noindent In order to interpret the oscillations of $G_{\mathrm m}$,
let us turn to the analytic model developed in
Sec.~\ref{sec:transport}. The height of the conductance peak follows
$\Delta_{1}^{\xi}(x_{0})$, given in Eq.~(\ref{eq:delta1}). It is
composed by the sum of two beating patterns, the terms
$g_{c,s}^{\xi}(x_{0})$, enveloped by the slow functions
$F_{c,s}^{\xi}(x_{0})$.
\begin{figure}[htbp]
\begin{center}
\includegraphics[width=8cm,keepaspectratio]{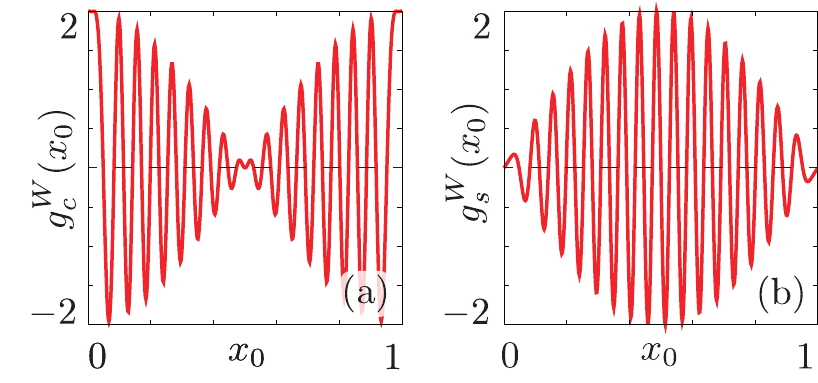}
\caption{(Color online) (a) Plot of $g_{c}^{W}(x_{0})$ as a function
  of $x_{0}$ (units $L$) for $N=16$ and $\alpha=k_{F}^{-1}=\pi/8L$; (b) Same as in
  (a) but for $g_{s}^{W}(x_{0})$.}
\label{fig:traverso8}
\end{center}
\end{figure}
The beatings arise from the superposition of the two cosine or sine
terms occurring in $g_{c,s}^{\xi}(x_{0})$ with a wavelength difference
of $2\pi/L$ and is therefore especially clear when $N\gg
1$. Figure~\ref{fig:traverso8} shows a typical example for the case of
the Wigner correction. The Friedel case is perfectly analogous, only
with half the wavelength.\\
\begin{figure}[htbp]
\begin{center}
\includegraphics[width=8cm,keepaspectratio]{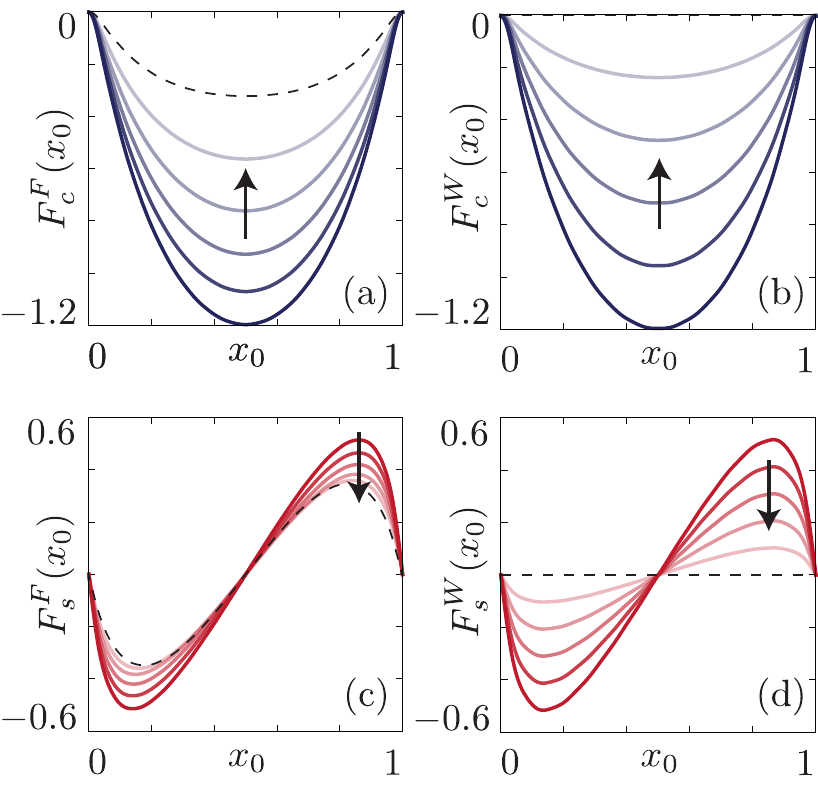}
\caption{(Color online) (a) Plot of $F_{c}^{F}(x_{0})$ (units
  $\varepsilon_{\sigma}^{-1}$) as a function of $x_{0}$ (units $L$)
  for different values of $g=1,\ 0.8,\ 0.6,\ 0.4,\ 0.2$ (from darker
  to lighter color). The arrow denotes the evolution of the family of
  curves as $g$ decreases (interaction strength increases); (b) Same
  as in (a) but for $F_{c}^{W}(x_{0})$; (c) Same as in (a) but for
  $F_{s}^{F}(x_{0})$; (c) Same as in (a) but for
  $F_{s}^{W}(x_{0})$. The dashed curve represents the asymptotic case
  $g\to 0$. All panels have been plotted for $N=16$.}
\label{fig:traverso9}
\end{center}
\end{figure}
\noindent The envelope functions $F_{c,s}^{\xi}(x_{0})$ parametrically
depend on $g$ and only very weakly on $N$ through the cutoff
$\alpha$. They are plotted in Fig.~\ref{fig:traverso9}. The overall shape
of the Friedel and Wigner cases is analogous. Important differences
however arise when the interaction strength increases ($g\to
0$). While $F_{c,s}^{F}(x_{0})$ tend asymptotically to a limit
function, the terms $F_{c,s}^{W}(x_{0})\to 0$. The mathematical origin
of this fact lies in the factor $\Lambda^{-1}$ into the summations
defining Eqns.~(\ref{eq:Fc},\ref{eq:Fs}). Indeed, see
Appendix~\ref{sec:appa}, for the Wigner case
$\Lambda\propto\varepsilon_{\rho}\propto g^{-1}$ and thus $F_{c,s}^{W}(x_{0})\to 0$
as $g\to 0$. On the other hand, for the Friedel case $\Lambda$ depends
both on $\varepsilon_{\rho}$ and $\varepsilon_{\sigma}$ and thus
non-vanishing contributions are still present even in the regime of
very strong interactions.\\
\noindent This confirms that the effects of the Friedel oscillations
on the conductance peak {\em amplitude} seem more robust again the
ones induced by Wigner oscillations.
\noindent The rather counter-intuitive conclusion that the corrections
to the maximum value of the linear conductance due to the Wigner term
vanish in the $g\to 0$ limit is very interesting. We want to remind
however that our discussion here concerns the {\em intrinsic}
dependence on the interaction parameter, regardless the weighting
factors of $\rho^{F,W}(x)$.
\section{Conclusions}
\label{sec:concl}
In this paper we have calculated the transport properties of an
interacting one-dimensional quantum dot, described by means of the
Luttinger liquid theory, in the presence of an AFM tip capacitively
coupled to it. We have considered both Friedel and Wigner corrections
to the electron density and evaluated their contributions both to the
chemical potential and to the tunneling rates to the lowest order in
the AFM-dot coupling.\\
\noindent To discuss the transport properties we have focused on the
linear regime, where we have shown that the AFM tip induces a shift of
the conductance peak and a renormalization of the conductance
strength, which has never been discussed in previous works on this
subject. Scanning the tip along the dot allows to observe oscillatory
patterns related to both the Friedel and the Wigner oscillations of
the electron density. A beating pattern emerges in the linear
conductance height oscillations. The Friedel and the Wigner
contributions show markedly distinct wavelengths allowing in principle
to distinguish the two effects. Surprisingly, we have found that as
the interaction strength grows, the effects of the Wigner oscillations
on the conductance height vanish, while those on the oscillations of
the peak position are reinforced.\\
\noindent Interesting effects are expected to show up even in the
nonlinear transport regime, also in connection to the possibility to
trigger and probe dot excited states. This will be the subject of
forthcoming investigations.\\

\noindent\textit{Acknowledgments.} Financial support by the EU- FP7
via ITN-2008-234970 NANOCTM is gratefully acknowledged.
\appendix
\section{Tunneling rate}
\label{sec:appa}
\noindent In this Appendix we outline the calculation of the tunneling
rates $\Gamma_{\mathcal{S}^{i}\to\mathcal{S}^{f}}^{\lambda}$
introduced in Sec.~\ref{sec:transport}, where
$|\mathcal{S}_{i,f}\rangle=|N_{\rho}^{i,f},N_{\sigma}^{i,f}\rangle$. \\
\noindent Let us start with the term $F_{0}^{\lambda}(t_{1},t_{2})$ in
Eq.~(\ref{eq:F0}). Following standard procedures already
outlined in literature~\cite{ioale1,ioale2} one obtains the standard result
\begin{equation}
F_{0}^{\lambda}(t_{1},t_{2})=\frac{\left|t_{\lambda}\right|^{2}}{2\pi\alpha}e^{-i\Delta E_{\lambda}(t_{2}-t_{1})}e^{-W_{\lambda}(t_{2}-t_{1})}e^{-W_{d}(t_{2}-t_{1})}\nonumber
\end{equation}
where
\begin{equation}
\Delta E_{\lambda}^{0}=\Delta n eV_{\lambda}+\mu_{0}(N_{\rho}^{i},N_{\sigma}^{i})
\end{equation}
with $\Delta n=(N_{\rho}^{f}-N_{\rho}^{i})$ and
$\mu_{0}(N_{\rho}^{i},N_{\sigma}^{i})$ is the bare chemical potential
of the dot, see Eq.~(\ref{eq:chempot0}).\\
\noindent Note that we have introduced the full chemical potential,
including tip-induced corrections. The kernel
\begin{equation}
e^{-W_{\lambda}(\tau)}=\langle\chi_{s,\lambda}^{\dagger}(x_{\lambda},\tau)\chi_{s,\lambda}(x_{\lambda},0)\rangle_{leads}\label{eq:Wl}
\end{equation}
is the correlation function for lead $\lambda$, which can be written as
\begin{equation}
e^{-W_{\lambda}(\tau)}=\nu_{\lambda}\int_{-\infty}^{\infty}\ dE\ e^{iE\tau}f(E)\, ,
\end{equation}
with $\nu_{\lambda}$ the density of states in lead $\lambda$ and
\begin{equation}
f(E)=\frac{1}{1+e^{\beta E}}
\end{equation}
is the Fermi function. In the following we will assume
$\nu_{\lambda}\equiv\nu_{0}$. Concerning the dot, $e^{-W_{d}(\tau)}$
stems from the thermal average over the collective excitations of the
initial states and sum over the final ones
with~\cite{ioale1,ioale2,iotobias}
\begin{equation}
W_{d}(\tau)=\frac{1}{2}\left[\frac{1}{g}W\left(\varepsilon_{\rho},\tau\right)+W\left(\varepsilon_{\sigma},\tau\right)\right]\, ,
\end{equation}
where
\begin{eqnarray}
W(\varepsilon,\tau)&=&\sum_{n>0}\frac{e^{-\pi\alpha n/L}}{n}\left[\coth{\left(\frac{\beta n\varepsilon}{2}\right)}\left(1-\cos(n\varepsilon\tau)\right)\right.\nonumber\\
&+&\left.i\sin(n\varepsilon\tau)\right]\, .\label{eq:W}
\end{eqnarray}
The perturbation terms $F_{1,2}^{(\lambda)}(t_{1},t_{2},t_{3})$ in
Eqns.~(\ref{eq:F1},\ref{eq:F2}) can be decomposed into Friedel
($\xi=F$) and Wigner ($\xi=W$) parts
\begin{equation}
F_{j}^{\lambda}(t_{1},t_{2},t_{3})=\sum_{\xi=F,W}F_{j}^{\xi,\lambda}(t_{1},t_{2},t_{3})\, ,
\end{equation}
with
\begin{widetext}
\begin{eqnarray}
F_{1}^{\xi,\lambda}(t_{1},t_{2},t_{3})&=&-i\frac{V_{\xi}}{\pi\alpha}K^{\xi}(x_{0})F_{0}^{\lambda}(t_{1},t_{3})\cos{\left[\mathcal{L}_{\xi}^{f}(X_{\lambda})+\Delta\mathcal{G}_{1}^{\xi,\lambda}(t_{1},t_{2},t_{3})\right]}\label{eq:f1}\\
F_{2}^{\xi,\lambda}(t_{1},t_{2},t_{3})&=&-i\frac{V_{\xi}}{\pi\alpha}K^{\xi}(x_{0})F_{0}^{\lambda}(t_{1},t_{3})\cos{\left[\mathcal{L}_{\xi}^{i}(X_{\lambda})+\Delta\mathcal{G}_{2}^{\xi,\lambda}(t_{1},t_{2},t_{3})\right]}\label{eq:f2}
\end{eqnarray}
\end{widetext}
where $X_{S}=x_{0}$, $X_{D}=L-x_{0}$, $K^{\xi}(x_{0})$ are defined in
Eqns.~(\ref{eq:Kf},\ref{eq:Kw}) and
\begin{eqnarray}
\mathcal{L}_{F}^{i/f}(x)&=&\Delta n\mathcal{L}(N_{s}^{i/f},x)\, ;\\
\mathcal{L}_{W}^{i/f}(x)&=&2\Delta n\mathcal{L}(N_{\rho}^{i/f}/2,x)\, .
\end{eqnarray}
The functions
\begin{equation}
-i\Delta\mathcal{G}_{j}^{\xi,\lambda}(t_{1},t_{2},t_{3})=(-1)^{j}\mathcal{G}_{\xi}^{\lambda}(t_{3}-t_{2})-\mathcal{G}_{\xi}^{\lambda}(t_{1+j}-t_{1})\, ,\nonumber
\end{equation}
with $j\in\{1,2\}$, are expressed in terms of
\begin{eqnarray}
  \mathcal{G}_{F}^{\lambda}(\tau)&=&-\frac{1}{2}\sum_{\nu=\rho,\sigma}\sum_{p=\pm1}pW\left(\varepsilon_{\nu},\tau+\frac{X_{\lambda}}{v_{\nu}}\right)\, ,\\
\mathcal{G}_{W}^{\lambda}(\tau)&=&-\sum_{p=\pm1}pW\left(\varepsilon_{\rho},\tau+\frac{X_{\lambda}}{v_{\rho}} \right)\, .
\end{eqnarray}
\\
\noindent Since $W(u)$ is periodic, it is convenient to exploit the Fourier series
\begin{equation}
e^{\pm\kappa W(u)}=\sum_{l=-\infty}^{\infty}b_{l}^{\pm,\kappa}e^{-ilu}\, .
\end{equation}
For $k_{B}T\ll\varepsilon_{\sigma}$ one can approximate $W_{l}$ with their expression~\cite{ioale1,ioale2} calculated for $T=0$,
\begin{eqnarray}
b_{l}^{+,\kappa}&=&\left(-e^{-\frac{l\pi\alpha}{L}}\right)^{l}\left(1-e^{-\frac{\alpha\pi}{L}}\right)^{\kappa}\frac{\Gamma(1+\kappa)\theta(l)}{l!\Gamma(1+\kappa-l)}\\
b_{l}^{-,\kappa}&=&\left(e^{-\frac{l\pi\alpha}{L}}\right)^{l}\left(1-e^{-\frac{\alpha\pi}{L}}\right)^{-\kappa}\frac{\Gamma(l+\kappa)}{l!\Gamma(\kappa)}\theta(l),
\end{eqnarray}
where $\Gamma(x)$ is the Euler gamma function. Inserting
Eq.~(\ref{eq:f1}) and Eq.~(\ref{eq:f2}) into Eq.~(\ref{eq:PP0}) and
Eq.~(\ref{eq:PP1}) it is possible to perform the time integration
exploiting the above Fourier expansions.\\
\noindent The tunneling rate in Eq.~(\ref{eq:pdotrate}) can be written
as
\begin{equation}
\Gamma_{\mathcal{S}^{i}\to\mathcal{S}^{f}}^{\lambda}=\Gamma_{\mathcal{S}^{i}\to\mathcal{S}^{f}}^{0,\lambda}+\delta\Gamma_{\mathcal{S}^{i}\to\mathcal{S}^{f}}^{F,\lambda}(x_{0})+\delta\Gamma_{\mathcal{S}^{i}\to\mathcal{S}^{f}}^{W,\lambda}(x_{0})\, ,\label{eq:rrategen}
\end{equation}
with
\begin{equation}
  \Gamma_{\mathcal{S}^{i}\to\mathcal{S}^{f}}^{0,\lambda}=\Gamma_{0}^{\lambda}\sum_{q_{\rho},q_{\sigma}}A_{q_{\rho},q_{\sigma}}f\left[\Delta E_{\lambda}^{0}+q_{\rho}\varepsilon_{\rho}+q_{\sigma}\varepsilon_{\sigma}\right]\label{eq:ratedellamamma}
\end{equation}
where $\Gamma_{0}^{\lambda}=\nu_{\lambda}\left|t_{\lambda}\right|^2/\pi\alpha$ and
\begin{equation}
A_{q_{\rho},q_{\sigma}}=b^{-,g/2}_{q_{\rho}}b^{-,1/2}_{q_{\sigma}}\, ;\label{eq:aqrho}
\end{equation}
the corrections induced by the tip read
\begin{widetext}
\begin{eqnarray}
\delta\Gamma_{\mathcal{S}^{i}\to\mathcal{S}^{f}}^{\xi,\lambda}(x_{0})&=&2V_{\xi}\Gamma_{0}^{\lambda}K^{\xi}(x_{0})\sum_{q_{\rho},q_{\sigma}}A_{q_{\rho},q_{\sigma}}R_{q_{\rho},q_{\sigma}}^{\xi,\lambda}(x_{0})\nonumber\\
&+&\left[\sum_{q_{\rho},q_{\sigma}}A_{q_{\rho},q_{\sigma}}\partial_{eV_\lambda}f(\Delta E_{\lambda}^{0}+q_{\rho}\varepsilon_{\rho}+q_{\sigma}\varepsilon_{\sigma})\right]V_{\xi}K^{\xi}(x_{0})\left\{\cos\left[\mathcal{L}_{\xi}^{f}(x_{0})\right]-\cos\left[\mathcal{L}_{\xi}^{i}(x_{0})\right]\right\}\label{eq:withcor}
\end{eqnarray}
\end{widetext}
where
\begin{widetext}
\begin{eqnarray}
\!\!\!\!\!\!\!\!\!\!R_{q_{\rho},q_{\sigma}}^{\xi,\lambda}(x_{0})&=&\left\{\sum_{\mathbf{n},\mathbf{m}|\Lambda\neq
  0}\frac{B^{\xi}_{\mathbf{n},\mathbf{m}}}{\Lambda}C_{\mathbf{n},\mathbf{m}}^{i,\xi}(X_{\lambda})+\sum_{\mathbf{n},\mathbf{m}}\bar{B}^{\xi}_{\mathbf{n},\mathbf{m}}C_{\mathbf{n},\mathbf{m}}^{f,\xi}(X_{\lambda})\left[\frac{1-\delta_{\bar{\Lambda},0}}{\bar{\Lambda}+0^{+}}+\frac{\delta_{\bar{\Lambda},0}}{2}\theta(\Lambda)\partial_{eV_\lambda}\right]\!\right\}F_{\mathbf{n}}(\Delta E_{\lambda}^{0},q_{\rho},q_{\sigma})\, ,\label{eq:TheRate}\\
\!\!\!\!\!\!\!\!F_{\mathbf{n}}(E,q_{\rho},q_{\sigma})&=&f\left[E+(q_{\rho}+n_{1}+n_{2})\varepsilon_{\rho}+(q_{\sigma}+n_{3}+n_{4})\varepsilon_{\sigma}\right]\, ,\label{eq:TheFF}
\end{eqnarray}
\end{widetext}
with $\mathbf{n}=\{n_{1},n_{2},n_{3},n_{4}\}$, $\mathbf{m}=\{m_{1},m_{2},m_{3},m_{4}\}$ and
\begin{equation}
C_{\mathbf{n},\mathbf{m}}^{i/f,\xi}(x)=\cos{\left[\mathcal{L}_{\xi}^{i/f}(x)+\left(k_{\mathbf{n}}+k_{\mathbf{m}}\right)x\right]}\, .
\end{equation}
Furthermore,
\begin{eqnarray}
\Lambda&=&\varepsilon_{\rho}(n_{1}+n_{2}+m_{1}+m_{2})+\varepsilon_{\sigma}(n_{3}+n_{4}+m_{3}+m_{4})\, ,\nonumber\\
\bar{\Lambda}&=&\varepsilon_{\rho}(m_{1}+m_{2}-n_{1}-n_{2})+\varepsilon_{\sigma}(m_{3}+m_{4}-n_{3}-n_{4})\, ,\nonumber
\end{eqnarray}
and $k_{\mathbf{n}}=\pi(n_{1}+n_{3}-n_{2}-n_{4})/L$ (analogous for
$k_{\mathbf{m}}$). Finally, the weights
$B_{\mathbf{n},\mathbf{m}}^{\xi}$ are
\begin{widetext}
\begin{eqnarray}
B^{F}_{\mathbf{n},\mathbf{m}}&=&b_{n_{1}}^{-,1/2}b_{n_{2}}^{+,1/2}b_{n_{3}}^{-,1/2}b_{n_{4}}^{+,1/2}b_{m_{1}}^{+,1/2}b_{m_{2}}^{-,1/2}b_{m_{3}}^{+,1/2}b_{m_{4}}^{-,1/2}\, ,\label{eq:Bf}\\
B^{W}_{\mathbf{n},\mathbf{m}}&=&b_{n_{1}}^{-,1}b_{n_{2}}^{+,1}b_{m_{1}}^{+,1}b_{m_{2}}^{-,1}\delta_{n_{3},0}\delta_{n_{4},0}\delta_{m_{3},0}\delta_{m_{4},0}\, ,\label{eq:Bw}
\end{eqnarray}
\end{widetext}
while $\bar{B}_{\mathbf{n},\mathbf{m}}^{\xi}$ is expressed in terms of $B_{\mathbf{n},\mathbf{m}}^{\xi}$ as
\begin{equation}
\bar{B}_{n_{1},n_{2},n_{3},n_{4},m_{1},m_{2},m_{3},m_{4}}^{\xi}=B_{n_{1},n_{2},n_{3},n_{4},m_{2},m_{1},m_{4},m_{3}}^{\xi}\, .\nonumber
\end{equation}\\

\noindent The second term in Eq.~(\ref{eq:withcor}) is easily
recognized to be the first-order term in $V_{\xi}$ of the expansion of
\begin{equation}
  \Gamma_{\mathcal{S}^{i}\to\mathcal{S}^{f}}^{0,\lambda}=\Gamma_{0}^{\lambda}\sum_{q_{\rho},q_{\sigma}}A_{q_{\rho},q_{\sigma}}f\left[\Delta E_{\lambda}+q_{\rho}\varepsilon_{\rho}+q_{\sigma}\varepsilon_{\sigma}\right]\label{eq:ratedellamamma}
\end{equation}
where
\begin{equation}
\Delta E_{\lambda}=\Delta n eV_{\lambda}+\mu_{d}(N_{\rho}^{i},N_{\sigma}^{i},x_{0})\label{eq:deltaefull}
\end{equation}
with $\mu_{d}(N_{\rho}^{i},N_{\sigma}^{i},x_{0})$, containing the
Friedel and Wigner corrections, defined in
Eq.~(\ref{eq:fullchempot}). From here on, we will always include in
both the bare rate, Eq.~(\ref{eq:ratedellamamma}) and in the explicit
corrections, Eq.~(\ref{eq:TheRate}), the full chemical potential
including its corrections replacing $\Delta E_{\lambda}^{0}$ with
$\Delta E_{\lambda}$.

\noindent At low temperatures, $\beta\varepsilon_{\sigma}\gg1$, useful
approximations to the tunneling rates when $\Delta E_{\lambda}\approx
0$ can be devised. This case is relevant for studying the linear
conductance peak.\\
\noindent Due to the argument of the Fermi function in
Eq.~(\ref{eq:ratedellamamma}), the only term not exponentially
suppressed is the one with $q_{\rho}=q_{\sigma}=0$. For the same
reason, in Eq.~(\ref{eq:TheRate}) the only terms which survive in the
summations composing $R_{q_{\rho},q_{\sigma}}^{\xi,\lambda}(x_{0})$
are those with $q_{\rho}=q_{\sigma}=0$ and $n_{1}=\ldots=n_{4}=0$. The
tunneling rate can be thus written as
\begin{equation}
\Gamma_{\mathcal{S}^{i}\to\mathcal{S}^{f}}^{\lambda}=\gamma^{\lambda}(x_{0})f(\Delta E_{\lambda})
\end{equation}
where
\begin{equation}
\gamma^{\lambda}(x_0)=\gamma^{0,\lambda}+\sum_{\xi=F,W}\delta\gamma^{\xi,\lambda}(x_{0})\, ,\label{eq:usefulcor}
\end{equation}
with
\begin{equation}
\gamma^{0,\lambda}=\Gamma_{0}^{\lambda}\left(1-e^{-\frac{\pi\alpha}{L}}\right)^{-\frac{1+g}{2}}
\end{equation}
and
\begin{widetext}
\begin{equation}
\delta\gamma^{\xi,\lambda}(x_{0})\approx  2\gamma^{0,\lambda}V_{\xi}K^{\xi}(x_{0})\sum_{\mathbf{m}\neq\mathbf{0}}\frac{1}{\Lambda}\left[B^{\xi}_{\mathbf{0},\mathbf{m}}C_{\mathbf{0},\mathbf{m}}^{i,\xi}(X_{\lambda})+\bar{B}^{\xi}_{\mathbf{0},\mathbf{m}}C_{\mathbf{0},\mathbf{m}}^{f,\xi}(X_{\lambda})\right]f(\Delta E_{\lambda})\, .
\end{equation}
\end{widetext}


\begin{thebibliography}{99}
\bibitem{comm} Other derivations of the density operator lead to a
  form which only slightly differs at the dot boundaries from the one
  we use.~\cite{haldane,sablikov}
\end{thebibliography}
\end{document}